\documentclass[twoside,12pt]{article}
\usepackage{exscale}                        
\usepackage[intlimits]{amsmath}       
\usepackage{amsfonts}
\usepackage{amssymb,amscd}
\usepackage{graphicx}
\usepackage{subfigure}

\newcommand{\be}{\begin{equation}}
\newcommand{\ee}{\end{equation}}
\newcommand{\bea}{\begin{eqnarray}}
\newcommand{\eea}{\end{eqnarray}}
\newcommand{\nn}{\nonumber}
\newcommand{\Eq}[1]{Eq.~(\ref{#1})}

\newcommand{\nf}{N_{f}\,}
\newcommand{\vf}{v_{F}\,}
\newcommand{\vd}{v_{\Delta}\,}
\newcommand{\qed}{$\mbox{QED}_3$}

\begin{document}
\title{ \vspace{1cm} Effects of Anisotropy in (2+1)-dimensional QED }
\author{J.~A.~Bonnet$^1$, C.~S.~Fischer$^{1,2}$, R.~Williams$^3$
\vspace*{3mm}\\
$^1$ Institut f\"ur Theoretische Physik, 
 Universit\"at Giessen, 35392 Giessen, Germany \\
$^2$GSI Helmholtzzentrum f\"ur Schwerionenforschung GmbH,\\ 
  Planckstr. 1  D-64291 Darmstadt, Germany \\
  $^3$ Dept. F\'isica Te\'orica I, Universidad Complutense, 28040 Madrid, Spain }
\maketitle
\begin{abstract}
\noindent
We summarize our results for the impact of anisotropic  
fermionic velocities in (2+1)-dimensional QED on the critical number 
of fermion flavors, $N^c_f$, and dynamical mass generation. 
We apply different approximation schemes for 
the gauge boson vacuum polarization and the fermion-boson vertex to analyze the according 
Dyson--Schwinger equations in a finite volume.
Our results point towards large variations of $N^c_f$ away from the isotropic point 
in agreement with other approaches. 
\end{abstract}

\section{Introduction}
Investigating strongly coupled theories has been a subject of great interest for many 
years. Amongst others, strong \qed~attracted a respectable amount of attention since 
it provides a comparably simple environment in which to study a variety of 
strong coupling phenomena. However, \qed~does not only serve as a test bed for investigations 
of more complicated theories. Viewed as a low-energy effective theory it
has applications to condensed matter systems such as high-temperature
superconductors (HTSs) or graphene. This strongly suggests that we need
to further elaborate its understanding.
In the case of HTSs, it was argued~\cite{Franz:2002qy,Herbut:2002yq} that the 
transition from the antiferromagnetic to the pseudogap phase can be described by \qed. 
The transition between both phases is then indicated by the generation of a dynamical
mass for the fermionic quasi-particles in the anti-ferromagnetic phase, whereas the 
pseudogap phase is chirally symmetric. The generation of such a mass term results 
from interactions of the fermionic quasi-particles with topological excitations that 
can be molded into the U(1) gauge fields of \qed. 
Experimental observations furthermore suggest the inclusion of anisotropic fermionic 
velocities in the theoretical description, which can be realized in the form of a 
constant, but non-trivial ``metric'' in the Lagrangian. While there have been 
extensive discussions on the critical number of fermion flavors for 
chiral symmetry breaking in isotropic \qed~both in the continuum 
\cite{Franz:2002qy,Herbut:2002yq,Appelquist:1988sr,Nash:1989xx,Maris:1996zg,Fischer:2004nq}
and at finite volume  
\cite{Goecke:2008zh,Hands:2002dv,Hands:2004bh,Strouthos:2008hs}, 
only few results are known which consider 
small~\cite{Lee:2002qza} or large anisotropies~\cite{Hands:2004ex,Thomas:2006bj,Concha:2009zj}. 
In this proceedings contribution, we summarize the results of our 
investigation of \qed~with (large) anisotropic velocities, obtained within the 
framework of Dyson--Schwinger equations on a compact manifold~\cite{Bonnet:2011hh}. 
First, we discuss 
some technical details concerning the underlying equations and introduce the 
approximation schemes that we use later on. In section \ref{sec:results} we 
present our results and compare them with findings within different approaches.

\section{Technical Details}
\subsection{The Dyson--Schwinger equations in anisotropic $\mbox{QED}_3$ 
\label{subsec:anisotropicdses}}
Motivated by the application of \qed~as a low-energy effective theory of high-temperature 
superconductors, we will investigate the anisotropic theory as
formulated and extensively discussed in Refs.~\cite{Franz:2002qy,Lee:2002qza,Hands:2004ex}.
Hence, we will here focus on the more important aspects of this
formulation of \qed.
We begin with the Lagrangian for anisotropic, (2+1)-dimensional QED containing $\nf$ fermion flavors 
in a four dimensional spinor representation 
\be
\mathcal{L}^{aniso}=\frac{N_f}{2}\sum_{j=1,2}
                         \bar{\Psi}_{j}
                         \left\{ \sum_{\mu=0}^{2}\gamma_{\nu}\sqrt{g}_{j,\nu\mu}
                                 \left(\partial_{\mu}+\mbox{i}\;a_{\mu}\right)
                         \right\}
                         \Psi_{j} .
\label{eq:lagrangian}
\ee
The fermionic spinors obey the Clifford algebra 
$\lbrace \gamma_{\mu},\gamma_{\nu}\rbrace=2\,\delta_{\mu\nu}$.\\
Furthermore, the Lagrangian \Eq{eq:lagrangian} contains the ``nodal'' sum  $\sum_{j=1,2}$ 
and the -- closely related -- metric-like factor $g_{j,\mu\nu}$. 
The sum arises from the HTS-inherent d-wave symmetry of the energy gap function, 
that leads to in total four zeroes  (the ``nodes'') that are found to lie on the Fermi surface.  
Two of these are related by symmetry and can be grouped in pairs which
are then
distinguished by their nodal index $j$. 
Fermionic excitations close to the zeroes of the energy gap function introduce the fermionic 
velocities $\vf$ and $\vd$ that are collected in the metric-like factor $g_{j,\mu\nu}$ 
defined as 

\vspace{5mm}
\begin{minipage}{.45\textwidth}
 \be
\left(g_{1}^{\mu\nu}\right)=\left(\begin{array}{ccc}
                                        1 &    0     & 0      \\
                                        0 & \,\vf^{2}& 0      \\
                                        0 &    0     & \vd^{2}
                                  \end{array}
                            \right)
\nn
\label{eq:metric 1}
\ee
\end{minipage}
 \begin{minipage}{.45\textwidth}
\be
\left(g_{2}^{\mu\nu}\right)=\left(\begin{array}{ccc}
                                         1&    0      &  0    \\
                                         0&  \vd^{2}  &  0    \\
                                         0&   0       &\vf^{2}
                                  \end{array}
                            \right).
\label{eq:metric 2}
\ee
\end{minipage}\\

\vspace{2mm}
\noindent
In the following, we will investigate the breaking of chiral symmetry that 
reduces the original  symmetry of the Lagrangian to
$\;U(2N_f)\rightarrow\;SU(N_f) \times SU(N_f) \times U(1) \times U(1)\;$,
when the fermions obtain dynamically generated mass terms.
The order parameter of the chiral phase transition  is given by the chiral condensate that 
can be determined from the trace of the fermionic propagator $S_{F,j}(\vec{p})$.

\begin{figure}[b!]
		  \hspace{3mm}
                  \begin{minipage}{0.45\textwidth}
                                            \hspace{3mm}
		                            \begin{center}
                                            \label{fig:dse}
                                            \includegraphics[width=\textwidth]{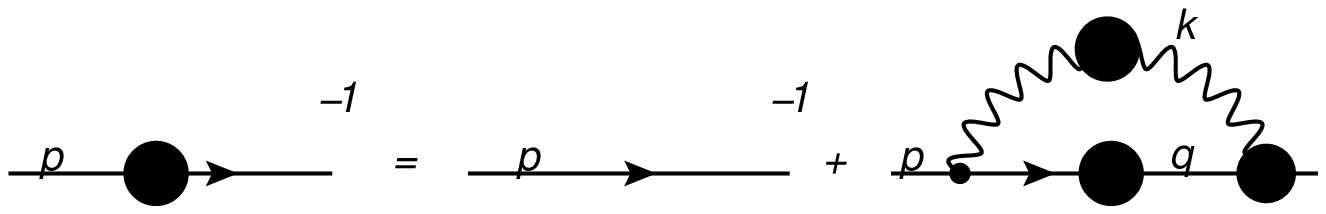}
                                            \end{center}
                \end{minipage}
                \hspace{5mm}
                \begin{minipage}{0.45\textwidth}
		                          \begin{center}
                                          
                                          \vspace*{11mm}
                                          \includegraphics[clip=true,trim= 0mm 0mm 3mm 0mm, width=\textwidth]{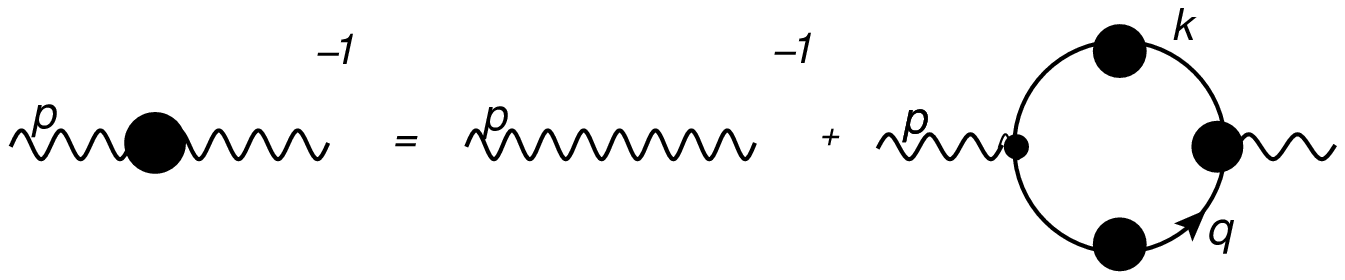}
                                          \end{center}
               \end{minipage}
                \caption{The diagrammatic representation of the Dyson--Schwinger equations
                         for the fermion (lhs) and gauge boson propagator (rhs).
                         Wiggly lines denote photon propagators, straight lines 
                         fermion propagators. 
                         A blob denotes a dressed propagator or vertex, whereas a 
                         dot stands for a bare fermion-photon vertex.}
\end{figure}

The diagrammatic representation of the corresponding Dyson--Schwinger equations 
for the fermion propagator $S_{F,i}$ and the bosonic propagator $D_{\mu\nu}$ 
are displayed in Fig.~\ref{fig:dse}. 
In Euclidean space-time, they are denoted as 
\bea
S_{F,i}^{-1}(\vec{p}\;)\;&=& S_{0}^{-1}(\,\vec{p}\,)
                          \;\;+\,Z_{1}\,e^{2}\int\frac{d^{3}q}{(2\pi)^{3}}
                                        (\sqrt{g}_{i,\mu\alpha}\gamma^{\alpha}\,S_{F,i}(\,\vec{q}\,)\,
                                         \sqrt{g}_{i,\nu\beta}\Gamma^{\beta}(\,\vec{q},\vec{p}\,)\,D_{\mu\nu}(\,\vec{k}\,)),
\label{eq:fermiondse} \\
D_{\mu\nu}^{-1}(\vec{p}\;)  &=& D_{0,\mu\nu}^{-1}(\,\vec{p}\,)
                               -Z_{1}\,e^{2}\,\frac{N_f}{2}\,\sum_{i=1,2}\int\frac{d^{3}q}{(2\pi)^{3}}
                                            \mbox{Tr}\left[
                                                           \sqrt{g}_{i,\mu\alpha}\gamma^{\alpha}\,S_{F,i}(\,\vec{q}\,)\;
                                                           \sqrt{g}_{i,\nu\beta}\Gamma^{\beta}(\,\vec{p},\vec{q}\,)\,S_{F,i}(\,\vec{k}\,))
                                                     \right],
\label{eq:photondse}
\eea
with the momentum $\,\vec{k}\,$ defined as the difference $\vec{p}-\vec{q}$ and $i=1,2$. 
Here, the renormalization constant $Z_{1}$ of the fermion-boson vertex,
 $\,\Gamma^{\beta}(\,\vec{p},\vec{q}\;)$, is included.

\noindent
We introduce several shorthands in order to keep track of the structure of the equations 
during further discussion. These are 
\begin{equation}
 \overline{p}_{i}^{2}:=p_{\mu}\;g_{i}^{\mu\nu}\;p_{\nu}
\label{eq:barconvention}
\end{equation}
and
\begin{align}
 \widetilde{p}_{\mu,i}:=A_{\mu,i}\left(\vec{p}\;\right) p_{\mu},\mbox{ (no summation convention !), }
\label{eq:tildeconvention}
\end{align}
where $A_{\mu,i}$ denotes the vectorial fermionic dressing function at node $i$.  \\
The anisotropic dressed propagators then are given as 
\bea
S_{F,i}^{-1}\left(\vec{p}\;\right)&=&B_{i}\left(\vec{p}\;\right)+\mbox{i}\;\sqrt{g_{i}}^{\mu\nu}\gamma_{\nu}\;\widetilde{p}_{\mu,i},
\label{eq:fermion}
\\
D_{\mu\nu}\left(\vec{p}\;\right)^{^{-1}}&=&p^{2}\left(\delta_{\mu\nu}-\frac{p_{\mu}p_{\nu}}{p^{2}}\right)
                                              +\Pi_{\mu\nu}\left(\vec{p}\;\right),
\label{eq:photon}
\eea
where $B_{i}$ denotes the scalar fermion dressing function at node \emph{i} and 
$\Pi_{\mu\nu}\left(\,\vec{p}\;\right)$ the vacuum polarization of the gauge boson field. 
As the free gauge bosons do not involve fermionic velocities, the bare propagator 
remains the isotropic one.\\
In order to solve the Dyson--Schwinger equations for the fermion and the gauge boson, 
we apply the minimal Ball-Chiu vertex construction as ansatz for the
fermion-gauge boson vertex. This is given by 
\be
	\Gamma^\beta_i(\vec{p},\vec{q}) = 
	\gamma^\beta \frac{A^\beta_i(\vec{p}) + A^\beta_i(\vec{q})}{2} 
	\label{vertex}
\ee
where no summation convention is used and $p,q$ are the fermion and anti-fermion momenta 
at the vertex.

\begin{figure}[b!]
		                            \begin{center}
                                            \label{fig:largen}
                                            \includegraphics[clip=true,trim= 0mm 25mm 60mm 145mm,width=0.9\textwidth]{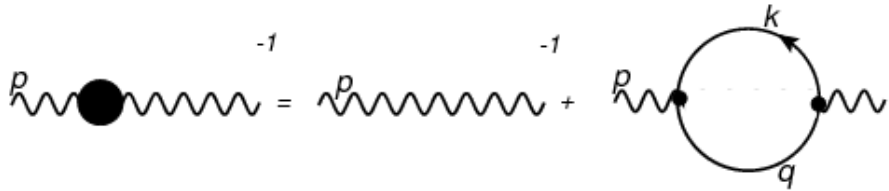}
                                            \end{center}
                \caption{The diagrammatic representation of the Dyson--Schwinger equation for the vacuumpolarization of
			 gauge bosons in leading order large-$\nf$ expansion.}
\end{figure}

\noindent
Furthermore, we approximate the gauge boson DSE in order to keep 
the equations manageable. We use two different approximation schemes. 
Firstly, we apply the large-$\nf$ approximation and secondly a more sophisticated model based
 on the results of Ref.~\cite{Fischer:2004nq}.\\
The large-$\nf$ approximation has been studied intensely for the limiting case of 
isotropic spacetime \cite{Appelquist:1988sr,Nash:1989xx} and also in an  expansion for 
small anisotropies (see Refs.~\cite{Franz:2002qy,Lee:2002qza}).
In this work, we are interested in the influence of large anisotropic  velocities 
and therefore make no further approximations concerning the ``metric''. 
Analyzing the large-$\nf$ approximation, we expect a qualitative estimate of 
the effects of anisotropy on the critical quantities for chiral phase transition. \\
The vacuum polarization in leading order large-$\nf$ expansion is given 
diagrammatically in Fig.~\ref{fig:largen}.

\noindent
After analytic integration (\cite{Franz:2002qy,Lee:2002qza}), the vacuum polarization is denoted as 
\bea
\Pi^{\mu\nu}\left(\vec{p}\;\right) & =&  \sum_{i}\sqrt{\overline{p}_i^2}
\left(g_{i}^{\mu\nu}-\frac{g_{i}^{\mu\alpha} p_{\alpha}\;g_{i}^{\nu\delta}p_{\delta}}
{\overline{p}_i^2}\right)\,\,
\Pi_i\left(\vec{p}\;\right).\nonumber\\
\Pi_i\left(\vec{p}\;\right) & = & 
\frac{e^{2}\, N_{f}}{16\,v_{F}v_{\Delta}} \frac{1}{\sqrt{\overline{p}_i^2}}.
\label{eq:largenvacuumpolarization}
\eea
Although the large-$\nf$ approximation is a good starting point 
to obtain first insights, it is well known that it contains 
several drawbacks. The origin of these lie in the fact that 
within a consistent large-$\nf$ approximation, the fermionic vector 
dressing function is independent of momentum, $A_{i,\mu} \equiv 1$. This momentum 
independence is in contrast to the power-law behaviour found in the infrared for the 
fermionic vector dressing function, the vacuum polarization and the vertex 
(see Ref.~\cite{Fischer:2004nq}). 
Moreover, the components of the vector dressing function provide 
for the renormalized fermionic anisotropic velocities ($c^R_s,v^R_{f}, v^R_{\Delta}$), 
which, in the general definition, would remain trivial for $A_{i,\mu} \equiv 1$. 
It is not intuitively clear if or how  their definition can be 
adjusted to be consistent with the large-$\nf$ approximation. 
Based on the results obtained in isotropic spacetime, 
we go beyond large-$\nf$ by applying a generalized gauge boson 
model that includes the anomalous dimension $\kappa$. 
We extracted the anomalous dimension in the isotropic limit with 
a value of $\kappa = 0.035$ and assume that we can neglect possible 
dependencies on the anisotropic velocities in this first and exploratory analysis. 
The improved vacuum polarization model reads
\be
	\Pi_i\left(\vec{p}\;\right) = \frac{e^2 N_f}{16\,v_{F}v_{\Delta}}  
	\left(\frac{1}{\sqrt{\overline{p}_i^2}}\frac{\overline{p}_i^2}{\overline{p}_i^2+e^2} 
	+ \frac{1}{\overline{p}_i^{1+2\kappa}}\frac{e^2}{{\overline{p}_i}^2+e^2}\right), \label{model}
\ee
with $\overline{{p}_i}^2$ defined in \Eq{eq:barconvention}.
With the approximation schemes at hand, we proceed to the fermionic DSEs 
by taking the appropriate traces of \Eq{eq:fermion}. We obtain 
\bea
%
B_{i}\left(\vec{p}\;\right) &      = & Z_{2} \; e^{2}\int\frac{d^{3}q}{\left(2\pi\right)^{3}}
                                       \,\, \frac{B_{i}\left(\vec{q}\;\right)g_{i}^{\mu\nu}D_{\mu\nu}(\vec{k}\;)}
                                             {B_{i}\left(\vec{q}\;\right)^{2}+(\overline{\widetilde{\vec{q}}}_{i}\;)^{2}}\,\,\,\frac{A_{\mu,i}(\vec{p}) + A_{\mu,i}(\vec{q})}{2},                                 
,
\label{eq:bdse}
                                 \\
%
A_{\mu,i}\left(\vec{p}\;\right) &    = & Z_{2}-\frac{Z_{2}\;e^{2}}{p_{\mu}}
                                        \int\frac{d^{3}q}{\left(2\pi\right)^{3}}
                                       \,\, 
                                            \frac{2\left(\widetilde{q}_{\lambda,i}\;g_{i}^{\lambda\nu}D_{\mu\nu}  (\vec{k}\;)\right)
                                                        -\widetilde{q}_{\mu,i}\;      g_{i}^{\lambda\nu}    D_{\lambda\nu}(\vec{k}\;)}
                                                 {B_{i}\left(\vec{q}\;\right)^{2}+(\;\overline{\widetilde{\vec{q}}}_{i}\;)^{2}}
                                        \,\, 
\frac{A_{\mu,i}(\vec{p}) + A_{\mu,i}(\vec{q})}{2},                                 
\eea
where there is again no summation convention for the index $\mu$.
Since the \emph{``metric''} factors are related by symmetry, the 
total number of equations to be solved can be reduced to eight 
DSEs. Nevertheless, the number of anisotropic equations is still more 
than double in comparison to isotropic spacetime (three equations), as we 
need to take the component-wise dressing of the fermion momentum 
and the more complex expression for the gauge boson vacuum polarization 
into account. 

In order to solve the Dyson-Schwinger equations self-consistently we
put the system in a box with finite volume $V=L^3$ and (anti-)periodic boundary 
conditions, i.e. we work on a three-torus in coordinate space which
translates to discrete momenta in momentum space
\cite{Goecke:2008zh,Fischer:2005ui}. The details of this 
procedure for our anisotropic case are given in Ref.\cite{Bonnet:2011hh}.

\section{Results \label{sec:results}}
In this section, we discuss our numerical results for the critical 
number of fermion flavours for chiral symmetry breaking within two 
approximation schemes: the large-$\nf$ approximation, including only 
the bare vertex, and secondly the gauge boson model including the 
minimal Ball-Chiu vertex construction. 
We determine the critical number of fermion flavors with help 
of the scalar fermion dressing function, evaluated at 
minimum momentum, which serves equally well as an order parameter as 
{\it e.g.} the chiral condensate. 

\subsection{\it The Large-N$_f$ Approximation \label{sec:largeN}}

\begin{figure}[h!]
		  \hspace{3mm}
\begin{center}
                  \begin{minipage}{0.45\textwidth}
                                            \hspace{3mm}
		                            \begin{center}
                                            \label{fig:phasediagram_largen}
                                            \includegraphics[width=\textwidth]{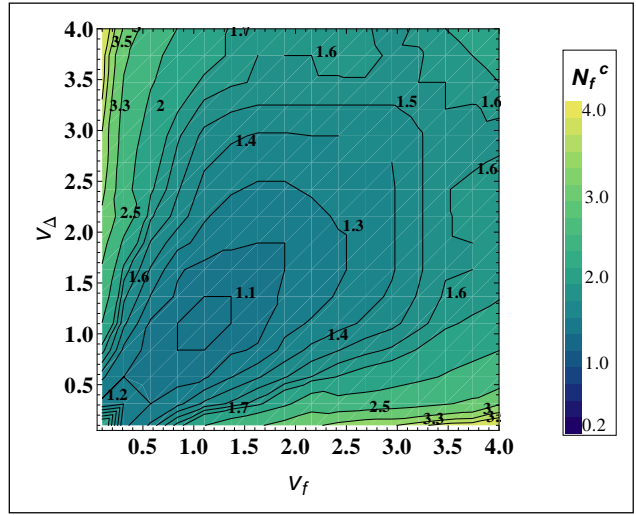}
                                            \end{center}
                \end{minipage}              
\end{center}
                \caption{Critical number of fermion flavours $N_f^c$ determined in the 
                                $1/N_f$-approximation and plotted as a function of the fermion velocities $(\vf,\vd)$.      
                             }
\end{figure}
\noindent
We obtained the results for the large-$\nf$ approximations 
on a torus with $Le^2 = 600 $ and $39^3$ momentum points, 
displayed in fig.~\ref{fig:phasediagram_largen}. The contour lines 
represent the values of $\nf$, supported by a corresponding 
color code. Notice that the contour lines should be considered 
to guide the eye, as the edges in some regions result from the 
sparseness of our grid. In principle, we expect smooth contours in 
$\vf-\vd-$space. 
For the isotropic point, ($\vf=\vd=1$), we find a value of $\nf^c \approx 1.0$.
As we expect this value to suffer from large volume effects \cite{Goecke:2008zh}, 
the deviation of about a factor of three from the continuum results 
is not surprising. 
Away from the isotropic point, we find increasing values for $\nf^c$, 
independent from increasing or decreasing $\vf$ and $\vd$, which 
is in agreement with earlier Dyson--Schwinger studies of large-$\nf$ 
approximation and an expansion for small anisotropies \cite{Lee:2002qza}. 
Within the large-$\nf$ approximation, these results remain valid also for 
larger anisotropic velocities. 
However, as argued in the previous section, this approximation 
suffers from several drawbacks, which we would like to circumvent 
by investigation of the gauge boson model. 

\subsection{\it The Gauge Boson Model}
%
\begin{figure}[h!]
                \hspace{5mm}
\begin{center}
                \begin{minipage}{0.45\textwidth}
		                          \begin{center}
                                          \label{fig:phasediagram_model}
                                          \vspace*{2mm}
                                          \includegraphics[clip=true,trim= 0mm 0mm 0mm 0mm, width=\textwidth]{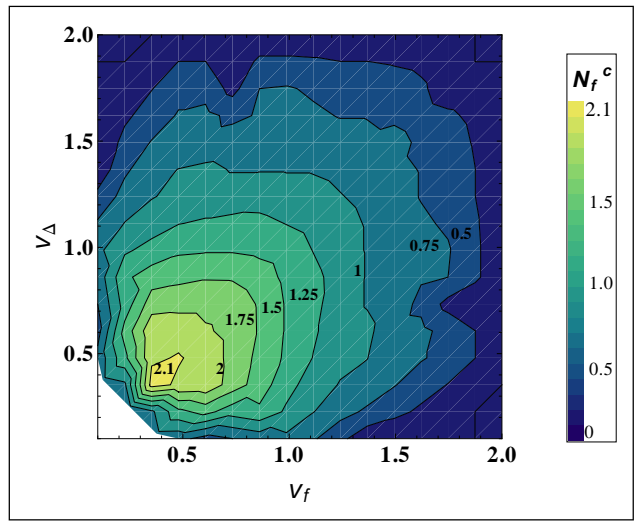}
                                          \end{center}
               \end{minipage}
\end{center}
                \caption{Critical number of fermion flavours $N_f^c$ determined from the
                               photon model \Eq{model} and the fermion-photon vertex \Eq{vertex} 
                               plotted as a function of the fermion velocities $(\vf,\vd)$.}
\end{figure}
%
\noindent
The results in the gauge boson model approximation scheme 
are obtained on a torus with $Le^2 = 600 $ and $39^3$ momentum points, too, and are 
displayed in fig.~\ref{fig:phasediagram_model}. 
The critical number of fermion flavors, $\nf^c$, is again indicated by contour lines 
and an additional color code. 
For the isotropic point, $\vf=\vd=1$ we find $\nf^{c,model}\approx 1.1$, which is slightly 
different from the value obtained from large-$\nf$ approximation. 
Away from the isotropic point, we find a completely different behaviour of 
$\nf^c$ within our gauge boson model. We find a maximum of $\nf^{c,model}$ around 
$(\vf,\vd)=(0.4,0.4)$ and a decreasing value of $\nf^{c,model}$ for smaller and larger 
anisotropies. While the critical number of fermion flavors never vanishes for $\vf,\vd < 1$, 
it reaches zero for $\vf,\vd \approx 2$, meaning that the theory is always in the chirally 
symmetric phase.  
Altogether, this behaviour is in accordance with findings in the
framework of a one-boson-exchange model 
\cite{Concha:2009zj}
and also with lattice calculations \cite{Hands:2004ex,Thomas:2006bj}.
These findings, with respect to the drawbacks of large-$\nf$ approximation, 
leads us to the conclusion that the results from large-$\nf$ approximation 
are not reliable as they dismiss important features of the theory influencing 
the chiral phasetransition of \qed.

\section{Conclusions}
We summarized our results of Ref.~\cite{Bonnet:2011hh} using a new model for 
the anisotropic gauge boson 
vacuum polarization including characteristics that are known to be important from 
isotropic continuum studies of \qed.  We discussed the results obtained within 
this model in context of large-$\nf$ calculations. We found a non-vanishing 
critical number of fermion flavours for fermionic velocities $\vf,\vd < 1$ and 
furthermore decreasing $\nf^{c,model}$ away from a maximum around 
$(\vf,\vd)\approx (0.4,0.4)$. These results agree qualitatively with findings 
from a continuum analysis of the one-boson exchange strength and with 
results from lattice calculations. We wish to emphasize that further 
investigations are necessary to quantitatively pin down the critical number
of fermion flavors as a function of the velocities. Since calculations on a
finite volume are necessarily hampered by large volume effects 
\cite{Goecke:2008zh,Gusynin:2003ww} it seems advisable to develop further a 
continuum approach to the problem. Ultimately, these efforts are bound to lead 
to a better understanding of \qed~as an effective theory of high temperature 
superconductors.

\section{Acknowledgements}
This work was supported by the Helmholtz-University Young Investigator Grant 
No. VH-NG-332 and by the Deutsche Forschungsgemeinschaft through SFB 634.


\end{document}